\documentclass[%
 aip,
 jap,
 onecolumn,
 amsfonts,
 amsmath,
 amssymb,
 superscriptaddress
 ]{revtex4-1}%

\usepackage{graphicx}
\usepackage{subfigure}
\usepackage{dcolumn}
\usepackage{bm}

\usepackage{SIunits}

\begin{document}

\title{Actively tunable slow light in a terahertz hybrid metal-graphene metamaterial}

\author{Tingting Liu}
\affiliation{Laboratory of Millimeter Wave and Terahertz Technology, School of Physics and Electronics Information, Hubei University of Education, Wuhan 430205, People's Republic of China}

\author{Chaobiao Zhou}
\affiliation{Wuhan National Laboratory for Optoelectronics, Huazhong University of Science and Technology, Wuhan 430074, People's Republic of China}

\author{Le Cheng}
\affiliation{Wuhan National Laboratory for Optoelectronics, Huazhong University of Science and Technology, Wuhan 430074, People's Republic of China}

\author{Xiaoyun Jiang}
\affiliation{Wuhan National Laboratory for Optoelectronics, Huazhong University of Science and Technology, Wuhan 430074, People's Republic of China}

\author{Guangzhao Wang}
\affiliation{School of Electronic Information Engineering, Yangtze Normal University, Chongqing 408100, People's Republic of China}

\author{Chen Xu}
\affiliation{Department of Physics, New Mexico State University, Las Cruces 88001, United State of America}

\author{Shuyuan Xiao}
\email{syxiao@hust.edu.cn}
\affiliation{Laboratory of Millimeter Wave and Terahertz Technology, School of Physics and Electronics Information, Hubei University of Education, Wuhan 430205, People's Republic of China}
\affiliation{Wuhan National Laboratory for Optoelectronics, Huazhong University of Science and Technology, Wuhan 430074, People's Republic of China}

\date{\today}

\begin{abstract}
We theoretically and numerically demonstrate an actively tunable slow light in a hybrid metal-graphene metamaterial in the terahertz (THz) regime. In the unit cell, the near field coupling between the metallic elements including the bright cut wire resonator and the dark double split-ring resonator gives rises to a pronounced transmission peak. By positioning a monolayer graphene under the dark mode resonator, an active modulation of the near field coupling is achieved via shifting the Fermi level of graphene. The physical origin can be attributed to the variation in the damping rate of the dark mode resonator arising from the conductive effect of graphene. Accompanied with the actively tunable near filed coupling effect is the dynamically controllable phase dispersion, allowing for the highly tunable slow light effect. This work offers an alternative way to design compact slow light devices in the THz regime for future optical signal processing applications.
\end{abstract}

\maketitle

\section{Introduction}\label{sec1}
The recent advances in artificially subwavelength-structured materials, i.e. metamaterials, have revolutionized the research fields for the efficient manipulation of electromagnetic waves.\cite{liu2011metamaterials,zheludev2012metamaterials} The unprecedented advantage of metamaterial lies in the exotic electromagnetic properties inaccessible in natural materials, which originates from the extremely geometrical scalability. This engineering tunability has inspired considerable works to fill the terahertz (THz) gap by designing proper metamaterial structures, realizing the interesting applications such as light modulators, perfect absorbers, and sensors.\cite{watts2014terahertz,min2016novel,watts2012metamaterial,wang2015novel,wang2016simple,cheng2016photoexcited,cheng2016infrared,meng2018simple} In particular, a wide variety of metamaterial structures composed of planar array of coupled resonators has been constructed to mimic the quantum phenomenon of electromagnetically induced transparency (EIT) in classical systems. In EIT metamaterial, the near field coupling between a bright and a dark mode resonator leads to a transmission window with steep dispersion, where the dark mode resonator showing sharp resonance is excited within the broad absorption band of the bright one. The large modification of the dispersive properties leads to the remarkable capability to slow down light, which shows promising prospects in optical buffering and storage for information processing.\cite{zhang2008plasmon,liu2009plasmonic,singh2011sharp,singh2009coupling,liu2012electromagnetically,zhu2013broadband,han2014engineering}

In practical implementations, it is highly desirable to actively control the slow light in the coupled resonator system of EIT metamaterial. Recent researchers have reported the active modulation schemes through dynamically controlling the photoconductive materials or thermal superconducting materials which are integrated in the coupled system.\cite{gu2012active,chatzakis2012reversible,roy2013ultrafast,cao2013plasmon,xu2016frequency,manjappa2017hybrid} For instance, by integrating photoconductive silicon into the metamaterial unit cell, Gu et al. experimentally demonstrated the optically tunable slow light and corresponding group delay for terahertz waves.\cite{gu2012active} Alternatively, graphene, a newly emerging material with distinguished electromagnetic properties, especially dynamically tunable surface conductivity, is proposed as an excellent competitor in actively manipulating slow light.\cite{zhao2016graphene,he2016terahertz,luo2016flexible,xia2016dynamically,fu2016dynamically,he2018graphene} As one typical example, Zhao et al. demonstrated the active modulation in a coupled system consisting of discrete monolayer graphene microstructure.\cite{zhao2016graphene} These approaches for tunable slow light are achieved by directly exciting the surface plasmon resonance in the structured graphene. However, the role of graphene as a conductive material is usually ignored and relatively few works proposed conductive graphene-based schemes and designs for active metamaterial. Very recently, some pioneering works have revealed the interaction between conductive graphene layer and the resonant THz metamaterial,\cite{li2016monolayer,xiao2017strong,chen2017study} which implies the great potentials of conductive graphene for the actively tunable slow light in a coupled system.

In this paper, we design a THz hybrid metal-graphene metamaterial structure for the dynamic modulation of slow light. In the coupled system, the bright and dark resonators are metallic and the monolayer graphene is deposited under the metal-based dark resonator in a continuous form. The near field coupling between the metallic resonators gives rise in a sharp transparency peak in the simulated transmission spectrum and an active modulation of the coupling effect can be achieved by shifting the Fermi level of graphene. The modulation mechanism is investigated with the electric field and surface charge distributions, as well as the classical coupled harmonic oscillator model. In particular, the actively tunable phase shift and group delay are achieved in the proposed structure, which would find applications in optical networks and THz wireless communications.

\section{Design and simulation of EIT structure}\label{sec2}
As shown in FIG. 1 (a) and (b), a hybrid metal-graphene metamaterial is designed to actively control slow light effect. The unit cell with a dimension of $P_x=100$ $\mu$m and $P_y=110$ $\mu$m is periodically patterned on the top of a silicon layers. Each unit cell of the proposed metamaterial consists of a cut wire resonator (CWR) and a double split-ring resonator (DSRR) both made of aluminum with a thickness of 200 nm, while the monolayer graphene is placed under the DSRR. The CWR has the length of $H=90$ $\mu$m and the width of $w=8$ $\mu$m, respectively. The DSRR has the base length of $L=60$ $\mu$m, the side length of $d=48$ $\mu$m, the width of $w=8$ $\mu$m, and the split gap length of $g=12$ $\mu$m. The DSRR is placed at a distance of $s=4$ $\mu$m away from the CWR.
\begin{figure}[htbp]
\centering
\includegraphics[scale=0.6]{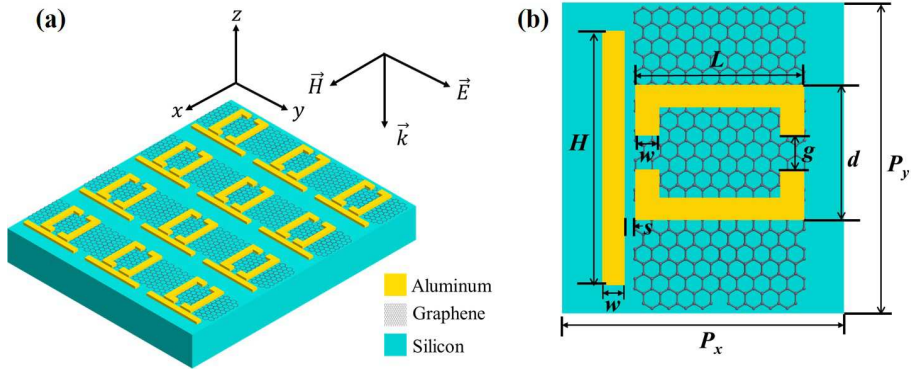}
\caption{\label{fig:1} (a) Three-dimensional schematic diagram of the hybrid metal-graphene EIT metamaterial.  (b) Schematic of the unit cell. Geometric parameters of the proposed structure: $P_x=100$ $\mu$m, $P_y=110$ $\mu$m, $H=90$ $\mu$m, $w=8$ $\mu$m, $s=4$ $\mu$m, $L=60$ $\mu$m, $d=48$ $\mu$m, $g=12$ $\mu$m. The incident electric field is along $y$ direction.}
\end{figure}

The numerical calculations are performed using the finite difference time domain (FDTD) method to investigate the near filed coupling effect of the hybrid metamaterial. The periodic boundary conditions are employed for the unit cell in $x$ and $y$ directions, and the perfect matched layer boundary condition for $z$ direction under the normal incidence. In simulations, the silicon substrate is assumed to be semi-infinite and its relative permittivity is taken as $\varepsilon_{Si}=3.42$. The permittivity of aluminum is described by Drude model $\varepsilon_{Al}=\varepsilon_{\infty}-\omega_p^2/(\omega^2+i\omega\gamma)$ with the plasmon frequency $\omega_p=2.24\times10^{16}$ rad/s and the damping constant $\gamma=1.22\times10^{14}$ rad/s, respectively.\cite{ordal1985optical} The optical properties of graphene can be characterized by the surface conductivity including the intraband and interband transition contributions. In THz regime, the intraband transition contribution is dominant and the conductivity can be modeled by Drude-like model $\sigma=ie^2E_F/[\pi\hbar^2(\omega+i\tau^{-1})]$ where $e$ is an electron charge, $E_f$ is the Fermi level of graphene, $\hbar$ is the reduced Planck constant.\cite{zhang2015towards,zhao2016tunable} The relaxation time $\tau$ can be calculated by $\tau=\mu E_F/(ev_F^2)$, where the carrier mobility $\mu=3000$ cm$^2/$V$\cdot$s and the Fermi velocity $v_F=1.1\times10^6$ m/s are employed in the simulations.\cite{jnawali2013observation,xiao2018active}

\section{Results and discussionss}\label{sec3}
To clarify the near field coupling effect in the proposed metamaterial, the transmission spectra of the isolated CWR array, the isolated DSRR array and the proposed structure array are calculated respectively, as shown in FIG. 2. When the incident electric field is along $y$ direction, the typical localized surface plasmon (LSP) resonance of the isolated CWR array is directly excited at the resonance frequency of 0.61 THz. Due to the structural symmetry with respect to the incident field, the isolated DSRR array is inactive with an inductive-capacitive (LC) resonance at the same frequency. Hence the CWR and the DSRR behave as the bright and dark resonators under the normal incidence, respectively. The bright and dark resonators show contrasting line widths within the frequency regime of interest, which fulfils the forming criterion of the EIT resonance. As a result, when the two types of resonators are arranged within a unit cell in a close proximity, a typical EIT resonance can be observed under $y$ polarized electric field excitation. As shown in FIG. 2, the destructive interference due to the near field coupling between them gives rise to a sharp transmission peak in a broad absorption regime.
\begin{figure}[htbp]
\centering
\includegraphics[scale=0.5]{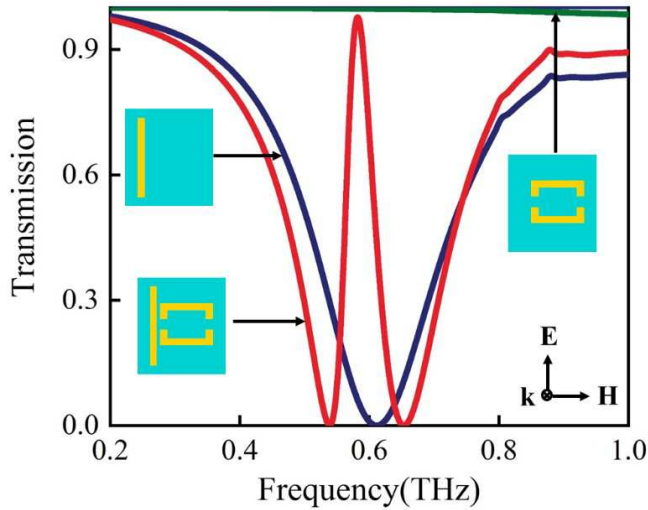}
\caption{\label{fig:2} Simulated transmission spectra of the isolated CWR array, the isolated DSRR array and the proposed structure array
}
\end{figure}

When the monolayer graphene is integrated into the metal-based structure, the near filed coupling effect would obtain an active modulation by shifting the Fermi level of graphene. Accordingly, the transmission peak of the proposed metamaterial would experience an on-to-off switching modulation, as FIG. 3(a) shows. For the case without graphene, a pronounced transmission peak can be observed between two resonance dips with the transmission amplitude of 97.18$\%$ at 0.58 THz. When the Fermi level of integrated graphene increases from 0.25 eV to 0.50 eV, the coupling effect between bright CWR and dark DSRR gradually weakens and the transmission peak undergoes a strong decline from 55.67$\%$ to 24$\%$. Upon further increase of the Fermi level, the coupling effect gradually diminishes. As the Fermi level increases to 1.00 eV, the transparency window disappear, showing a broad LSP resonance dip with a low transmission amplitude of only 0.59$\%$ in the spectra. Therefore, by shifting the Fermi level of graphene, an on-to-off switch of transmission peak associated with an active modulation of the near field coupling is achieved in the proposed metamaterial.
\begin{figure}[htbp]
\centering
\includegraphics[scale=0.6]{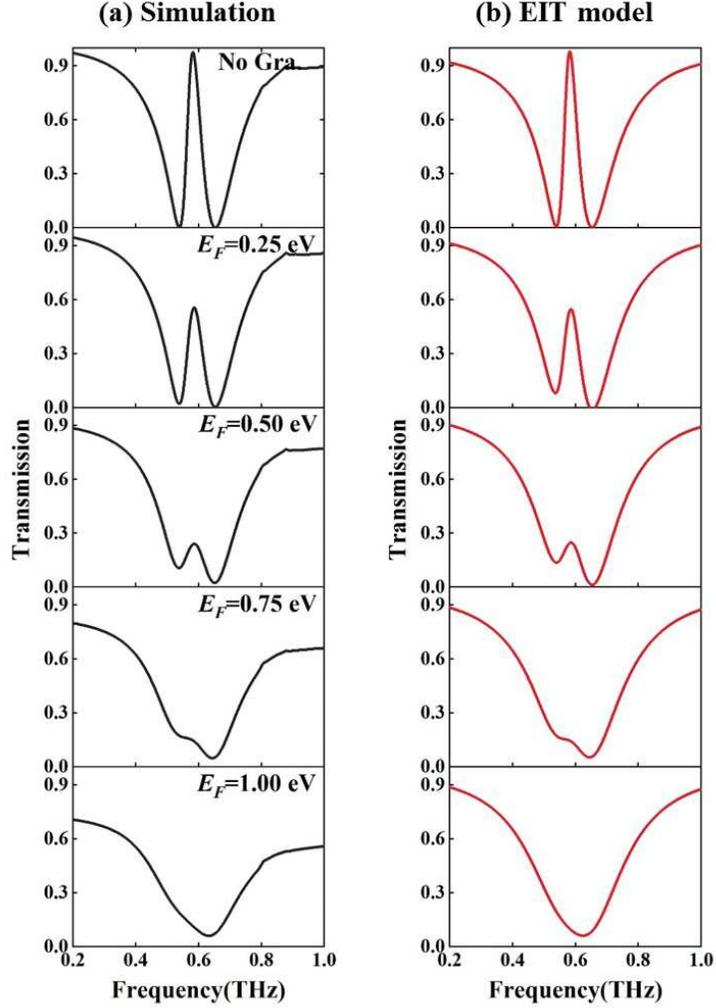}
\caption{\label{fig:3}The transmission spectra of the proposed metamaterial at different Fermi levels of graphene. (a) The simulated results. (b) The theoretical fitting spectra.}
\end{figure}

To elucidate the physical mechanism of the active modulation, the classical coupled harmonic oscillator model is adopted to analyze the coupling effect between the bright and dark resonators in the proposed EIT metamaterial. The bright CWR in the unit cell is represented by oscillator 1, which is directly excited by the incident electromagnetic field $E$. The dark DSRR is represented by oscillator 2, which can be indirectly excited via the near field coupling between the two oscillators. For the proposed metamaterial, the interaction between the two resonators can be analytically described by the differential equations as follows\cite{liu2009plasmonic}
\begin{equation}
    \ddot{x}_1+\gamma_1 \dot{x}_1+\omega_0^2 x_1+\kappa x_2=E,\label{eq1}
\end{equation}
\begin{equation}
    \ddot{x}_2+\gamma_2\dot{x}_2+\omega_1^2 x_2+\kappa x_1=0,\label{eq2}
\end{equation}
where $x_1$, $x_2$, $\gamma_1$ and $\gamma_2$ are the resonance amplitudes and the damping rates of the bright and dark modes, respectively. $\omega_0$ and $\omega_0+\delta$ are the resonance frequencies of bright and dark modes, respectively. $\kappa$ is the coupling coefficient between the two resonators, and $E$ represents the incident electric field. By solving the equations in (1) and (2) with the approximation of $\omega-\omega_0<<\omega_0$, the susceptibility $\chi$ of the unit cell can be expressed by\cite{luo2016flexible}
\begin{equation}
    \chi=\chi_r+i\chi_i\propto\frac{\omega-\omega_1+i\frac{\gamma_2}{2}}{(\omega-\omega_0+i\frac{\gamma_1}{2})(\omega-\omega_1+i\frac{\gamma_2}{2})-\frac{\kappa^2}{4}}.\label{eq3}
\end{equation}
Since the energy dissipation is proportional to the imaginary part $\chi_i$, the transmission T can be obtained as $T=1-g\chi_i$, where $g$ describes the coupling strength of the bright mode with the incident electric field $E$.

Based on the coupled harmonic oscillator model, the theoretical fitted transmission spectra at different Fermi levels of graphene are depicted in FIG. 3(b). It is observed that the fitted spectra show an excellent agreement with the simulated results. On the other hand, the variations of the fitting parameters as the Fermi level increases are illustrated in FIG. 4. During the modulation process, the fitting parameters of $\gamma_1$, $\kappa$, $\delta$ keeps basically constant, however, the damping rate of the dark DSRR $\gamma_2$ increases by two orders of magnitude from 0.005 to 0.12 THz. Thus, it can be concluded that the modulation of coupling effect is attributed to the increase in the damping rate of the dark resonator. The monolayer graphene placed under the dark DSRR behaves as a metal, and connects the two ends of each split with the high conductivity. Hence the conductive graphene enhances the loss in the dark mode and weakens the destructive interference between bright and mode modes. When the Fermi level reaches 1.00 eV, the large damping rate of dark mode $\gamma_2$ implies that the losses becomes too significant to maintain the dark resonance mode of DSRR, giving rise to the complete disappearance of the transmission peak.
\begin{figure}[htbp]
\centering
\includegraphics[scale=0.5]{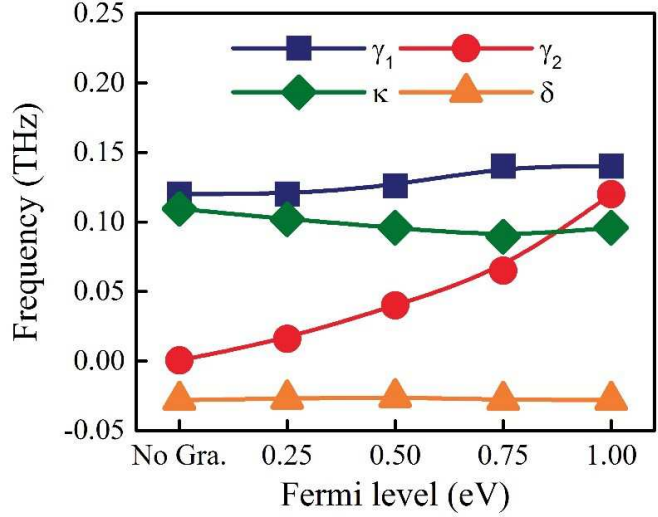}
\caption{\label{fig:4}The variations in the fitting parameters of the coupled harmonic oscillator model at different Fermi levels of graphene.}
\end{figure}

To further investigate the physical origin for the active modulation of the near filed coupling in the metamaterial, the distributions of the electric field and surface charge with the change in the Fermi level are provided in FIG. 5. For the case without graphene, the LC resonance mode of DSRR is excited through the near field coupling with the bright CWR and then destructively interferes with the LSP mode of CWR. This can be reflected by the field distributions in FIG. 5(a) and (f). Both the strong electric field enhancement and the accumulated opposite charges are centered around the two ends of each split of DSRR. In this case, the dark DSRR has a very small damping rate, indicating the small loss and the strong excitation of the dark resonance mode. When integrated under the dark DSRR, the monolayer graphene as a conductive layer connects the two ends of each split and its surface conductivity would increase as Fermi level becomes larger. Hence, the opposite charges are recombined and neutralized through this conductive layer, which hampers the excitation of the dark DSRR. When the Fermi level is 0.50 eV, the electric field enhancement and the opposite charges in the DSRR show a remarkable decline, while the field distributions in CWR start to strengthen, as shown in FIG.5 (c) and (h). With the Fermi level of 1.00 eV, the conductive graphene completely connects the two ends of each split of DSRR, strongly suppressing the near filed coupling between the two resonators. The distributions of the electric field and the surface charge of the dark DSRR are completely eliminated, while the field distributions are focused around the bright CWR with LSP resonance in FIF. 5 (e) and (j). The dark DSRR exhibits too large damping rate to sustain the LC resonance, leading to the disappearance of the coupling effect between the two resonance modes. Therefore, the physical origin of active modulation of the coupling effect lies in the change in the damping rate of the dark DSRR due to the recombination effect of the conductive graphene.
\begin{figure}[htbp]
\centering
\includegraphics[scale=0.7]{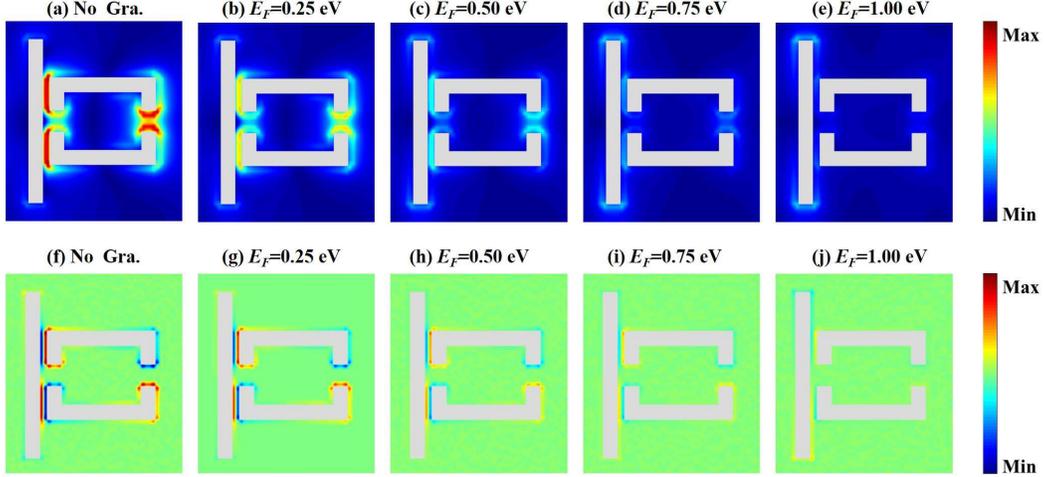}
\caption{\label{fig:5}The variations of the distributions of (a)-(e) the electric field and (f)-(j) the surface charge at different Fermi levels of graphene at corresponding resonance frequencies in the proposed metamaterial.}
\end{figure}

The actively controllable near field coupling of EIT resonance leads to the tunable slow light in the proposed metamaterial. The capability to slow down the speed of light can be described by the group delays, $\tau_g={d\psi}/{d\omega}$, where $\psi$ represents the phase shift of the transmission.\cite{lu2012plasmonic} It is evident that both the phase shift and the group delay experience a great modulation as the Fermi level of graphene increases in FIG. 6(a) and (b). For the initial case without graphene, the phase shows a steepest shift inside the transparency window and the maximum group delay is calculated as 6.74 ps. When the monolayer graphene is integrated, the slow light capability of the EIT metamaterial gradually weakens with the increasing Fermi level. For example, group delay decreases from 4.20 ps to 1.61 ps as Fermi level increases from 0.25 eV to 0.5 eV. When Fermi level reaches the maximum value of 1.00 eV, the phase shift and the group delay disappear within the transparency window. The metamaterial gradually loses its slow light capability due to weakening near field coupling effect.
\begin{figure}[htbp]
\centering
\includegraphics[scale=0.6]{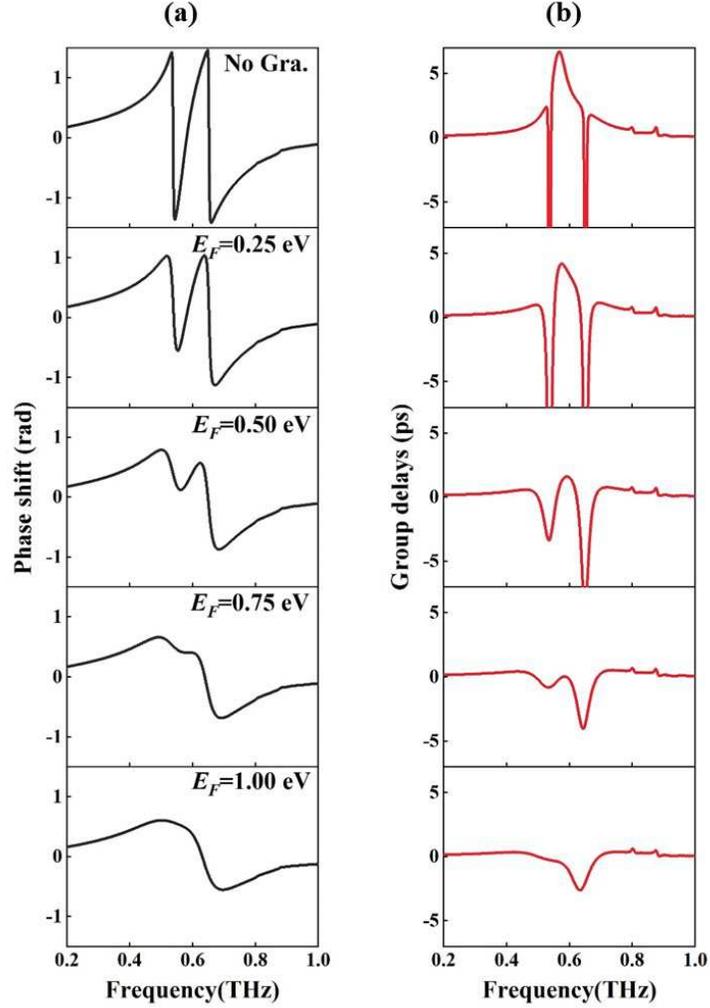}
\caption{\label{fig:6}The variations in (a) the phase shift and (b) the group delay at different Fermi levels of graphene.}
\end{figure}

\section{Conclusion}\label{sec4}
In conclusion, we have theoretically demonstrated an active modulation of slow light associated with the near filed coupling effect. By integrating the monolayer graphene into the THz metamaterial composed of metallic bright CWR and the dark DSRR, the near field coupling effect between the bright and dark resonators is dynamically controlled via shifting the Fermi level of graphene placed the dark DSRR. Based on the coupled harmonic oscillator model and the distributions of electric filed and surface charge, the theoretical analysis reveals that the underlying physical mechanism can be attributed to the increasing damping rate of the dark DSRR due to recombination effect of the conductive graphene. During the modulation process, the actively tunable group delays in the proposed EIT metamaterial are demonstrated for the slow light effect. The proposed hybrid metal-graphene metamaterial provides an alternative way to realize tunable slow light effect, and offers avenues to design highly compact THz functional devices for signal processing applications.

\begin{acknowledgments}
This work is supported by the National Natural Science Foundation of China (Grant No. 61775064), the Fundamental Research Funds for the Central Universities (HUST: 2016YXMS024) and the Natural Science Foundation of Hubei Province (Grant No. 2015CFB398 and 2015CFB502).
\end{acknowledgments}

\nocite{*}

\begin{thebibliography}{38}%
\makeatletter
\providecommand \@ifxundefined [1]{%
 \@ifx{#1\undefined}
}%
\providecommand \@ifnum [1]{%
 \ifnum #1\expandafter \@firstoftwo
 \else \expandafter \@secondoftwo
 \fi
}%
\providecommand \@ifx [1]{%
 \ifx #1\expandafter \@firstoftwo
 \else \expandafter \@secondoftwo
 \fi
}%
\providecommand \natexlab [1]{#1}%
\providecommand \enquote  [1]{``#1''}%
\providecommand \bibnamefont  [1]{#1}%
\providecommand \bibfnamefont [1]{#1}%
\providecommand \citenamefont [1]{#1}%
\providecommand \href@noop [0]{\@secondoftwo}%
\providecommand \href [0]{\begingroup \@sanitize@url \@href}%
\providecommand \@href[1]{\@@startlink{#1}\@@href}%
\providecommand \@@href[1]{\endgroup#1\@@endlink}%
\providecommand \@sanitize@url [0]{\catcode `\\12\catcode `\$12\catcode
  `\&12\catcode `\#12\catcode `\^12\catcode `\_12\catcode `\%12\relax}%
\providecommand \@@startlink[1]{}%
\providecommand \@@endlink[0]{}%
\providecommand \url  [0]{\begingroup\@sanitize@url \@url }%
\providecommand \@url [1]{\endgroup\@href {#1}{\urlprefix }}%
\providecommand \urlprefix  [0]{URL }%
\providecommand \Eprint [0]{\href }%
\providecommand \doibase [0]{http://dx.doi.org/}%
\providecommand \selectlanguage [0]{\@gobble}%
\providecommand \bibinfo  [0]{\@secondoftwo}%
\providecommand \bibfield  [0]{\@secondoftwo}%
\providecommand \translation [1]{[#1]}%
\providecommand \BibitemOpen [0]{}%
\providecommand \bibitemStop [0]{}%
\providecommand \bibitemNoStop [0]{.\EOS\space}%
\providecommand \EOS [0]{\spacefactor3000\relax}%
\providecommand \BibitemShut  [1]{\csname bibitem#1\endcsname}%
\let\auto@bib@innerbib\@empty
\bibitem [{\citenamefont {Liu}\ and\ \citenamefont
  {Zhang}(2011)}]{liu2011metamaterials}%
  \BibitemOpen
  \bibfield  {author} {\bibinfo {author} {\bibfnamefont {Y.}~\bibnamefont
  {Liu}}\ and\ \bibinfo {author} {\bibfnamefont {X.}~\bibnamefont {Zhang}},\
  }\href@noop {} {\bibfield  {journal} {\bibinfo  {journal} {Chem. Soc. Rev.}\
  }\textbf {\bibinfo {volume} {40}},\ \bibinfo {pages} {2494} (\bibinfo {year}
  {2011})}\BibitemShut {NoStop}%
\bibitem [{\citenamefont {Zheludev}\ and\ \citenamefont
  {Kivshar}(2012)}]{zheludev2012metamaterials}%
  \BibitemOpen
  \bibfield  {author} {\bibinfo {author} {\bibfnamefont {N.~I.}\ \bibnamefont
  {Zheludev}}\ and\ \bibinfo {author} {\bibfnamefont {Y.~S.}\ \bibnamefont
  {Kivshar}},\ }\href@noop {} {\bibfield  {journal} {\bibinfo  {journal} {Nat.
  Mater.}\ }\textbf {\bibinfo {volume} {11}},\ \bibinfo {pages} {917} (\bibinfo
  {year} {2012})}\BibitemShut {NoStop}%
\bibitem [{\citenamefont {Watts}\ \emph {et~al.}(2014)\citenamefont {Watts},
  \citenamefont {Shrekenhamer}, \citenamefont {Montoya}, \citenamefont
  {Lipworth}, \citenamefont {Hunt}, \citenamefont {Sleasman}, \citenamefont
  {Krishna}, \citenamefont {Smith},\ and\ \citenamefont
  {Padilla}}]{watts2014terahertz}%
  \BibitemOpen
  \bibfield  {author} {\bibinfo {author} {\bibfnamefont {C.~M.}\ \bibnamefont
  {Watts}}, \bibinfo {author} {\bibfnamefont {D.}~\bibnamefont {Shrekenhamer}},
  \bibinfo {author} {\bibfnamefont {J.}~\bibnamefont {Montoya}}, \bibinfo
  {author} {\bibfnamefont {G.}~\bibnamefont {Lipworth}}, \bibinfo {author}
  {\bibfnamefont {J.}~\bibnamefont {Hunt}}, \bibinfo {author} {\bibfnamefont
  {T.}~\bibnamefont {Sleasman}}, \bibinfo {author} {\bibfnamefont
  {S.}~\bibnamefont {Krishna}}, \bibinfo {author} {\bibfnamefont {D.~R.}\
  \bibnamefont {Smith}}, \ and\ \bibinfo {author} {\bibfnamefont {W.~J.}\
  \bibnamefont {Padilla}},\ }\href@noop {} {\bibfield  {journal} {\bibinfo
  {journal} {Nat. Photonics}\ }\textbf {\bibinfo {volume} {8}},\ \bibinfo
  {pages} {605} (\bibinfo {year} {2014})}\BibitemShut {NoStop}%
\bibitem [{\citenamefont {Min}\ \emph {et~al.}(2016)\citenamefont {Min},
  \citenamefont {Sun}, \citenamefont {Zhang}, \citenamefont {Ding},
  \citenamefont {Shen},\ and\ \citenamefont {Sun}}]{min2016novel}%
  \BibitemOpen
  \bibfield  {author} {\bibinfo {author} {\bibfnamefont {W.}~\bibnamefont
  {Min}}, \bibinfo {author} {\bibfnamefont {H.}~\bibnamefont {Sun}}, \bibinfo
  {author} {\bibfnamefont {Q.}~\bibnamefont {Zhang}}, \bibinfo {author}
  {\bibfnamefont {H.}~\bibnamefont {Ding}}, \bibinfo {author} {\bibfnamefont
  {W.}~\bibnamefont {Shen}}, \ and\ \bibinfo {author} {\bibfnamefont
  {X.}~\bibnamefont {Sun}},\ }\href@noop {} {\bibfield  {journal} {\bibinfo
  {journal} {J. Opt.}\ }\textbf {\bibinfo {volume} {18}},\ \bibinfo {pages}
  {065103} (\bibinfo {year} {2016})}\BibitemShut {NoStop}%
\bibitem [{\citenamefont {Watts}, \citenamefont {Liu},\ and\ \citenamefont
  {Padilla}(2012)}]{watts2012metamaterial}%
  \BibitemOpen
  \bibfield  {author} {\bibinfo {author} {\bibfnamefont {C.~M.}\ \bibnamefont
  {Watts}}, \bibinfo {author} {\bibfnamefont {X.}~\bibnamefont {Liu}}, \ and\
  \bibinfo {author} {\bibfnamefont {W.~J.}\ \bibnamefont {Padilla}},\
  }\href@noop {} {\bibfield  {journal} {\bibinfo  {journal} {Adv. Mater.}\
  }\textbf {\bibinfo {volume} {24}} (\bibinfo {year} {2012})}\BibitemShut
  {NoStop}%
\bibitem [{\citenamefont {Wang}\ \emph {et~al.}(2015)\citenamefont {Wang},
  \citenamefont {Zhai}, \citenamefont {Wang}, \citenamefont {Huang},\ and\
  \citenamefont {Wang}}]{wang2015novel}%
  \BibitemOpen
  \bibfield  {author} {\bibinfo {author} {\bibfnamefont {B.-X.}\ \bibnamefont
  {Wang}}, \bibinfo {author} {\bibfnamefont {X.}~\bibnamefont {Zhai}}, \bibinfo
  {author} {\bibfnamefont {G.-Z.}\ \bibnamefont {Wang}}, \bibinfo {author}
  {\bibfnamefont {W.-Q.}\ \bibnamefont {Huang}}, \ and\ \bibinfo {author}
  {\bibfnamefont {L.-L.}\ \bibnamefont {Wang}},\ }\href@noop {} {\bibfield
  {journal} {\bibinfo  {journal} {J. Appl. Phys.}\ }\textbf {\bibinfo {volume}
  {117}},\ \bibinfo {pages} {014504} (\bibinfo {year} {2015})}\BibitemShut
  {NoStop}%
\bibitem [{\citenamefont {Wang}, \citenamefont {Wang},\ and\ \citenamefont
  {Sang}(2016)}]{wang2016simple}%
  \BibitemOpen
  \bibfield  {author} {\bibinfo {author} {\bibfnamefont {B.-X.}\ \bibnamefont
  {Wang}}, \bibinfo {author} {\bibfnamefont {G.-Z.}\ \bibnamefont {Wang}}, \
  and\ \bibinfo {author} {\bibfnamefont {T.}~\bibnamefont {Sang}},\ }\href@noop
  {} {\bibfield  {journal} {\bibinfo  {journal} {J. Phys. D: Appl. Phys.}\
  }\textbf {\bibinfo {volume} {49}},\ \bibinfo {pages} {165307} (\bibinfo
  {year} {2016})}\BibitemShut {NoStop}%
\bibitem [{\citenamefont {Cheng}, \citenamefont {Gong},\ and\ \citenamefont
  {Cheng}(2016)}]{cheng2016photoexcited}%
  \BibitemOpen
  \bibfield  {author} {\bibinfo {author} {\bibfnamefont {Y.}~\bibnamefont
  {Cheng}}, \bibinfo {author} {\bibfnamefont {R.}~\bibnamefont {Gong}}, \ and\
  \bibinfo {author} {\bibfnamefont {Z.}~\bibnamefont {Cheng}},\ }\href@noop {}
  {\bibfield  {journal} {\bibinfo  {journal} {Opt. Commun.}\ }\textbf {\bibinfo
  {volume} {361}},\ \bibinfo {pages} {41} (\bibinfo {year} {2016})}\BibitemShut
  {NoStop}%
\bibitem [{\citenamefont {Cheng}\ \emph {et~al.}(2016)\citenamefont {Cheng},
  \citenamefont {Mao}, \citenamefont {Wu}, \citenamefont {Wu},\ and\
  \citenamefont {Gong}}]{cheng2016infrared}%
  \BibitemOpen
  \bibfield  {author} {\bibinfo {author} {\bibfnamefont {Y.}~\bibnamefont
  {Cheng}}, \bibinfo {author} {\bibfnamefont {X.~S.}\ \bibnamefont {Mao}},
  \bibinfo {author} {\bibfnamefont {C.}~\bibnamefont {Wu}}, \bibinfo {author}
  {\bibfnamefont {L.}~\bibnamefont {Wu}}, \ and\ \bibinfo {author}
  {\bibfnamefont {R.}~\bibnamefont {Gong}},\ }\href@noop {} {\bibfield
  {journal} {\bibinfo  {journal} {Opt. Mater.}\ }\textbf {\bibinfo {volume}
  {53}},\ \bibinfo {pages} {195} (\bibinfo {year} {2016})}\BibitemShut
  {NoStop}%
\bibitem [{\citenamefont {Meng}\ \emph {et~al.}(2018)\citenamefont {Meng},
  \citenamefont {Wang}, \citenamefont {Zhai}, \citenamefont {Liu},\ and\
  \citenamefont {Xia}}]{meng2018simple}%
  \BibitemOpen
  \bibfield  {author} {\bibinfo {author} {\bibfnamefont {H.-Y.}\ \bibnamefont
  {Meng}}, \bibinfo {author} {\bibfnamefont {L.-L.}\ \bibnamefont {Wang}},
  \bibinfo {author} {\bibfnamefont {X.}~\bibnamefont {Zhai}}, \bibinfo {author}
  {\bibfnamefont {G.-D.}\ \bibnamefont {Liu}}, \ and\ \bibinfo {author}
  {\bibfnamefont {S.-X.}\ \bibnamefont {Xia}},\ }\href@noop {} {\bibfield
  {journal} {\bibinfo  {journal} {Plasmonics}\ }\textbf {\bibinfo {volume}
  {13}},\ \bibinfo {pages} {269} (\bibinfo {year} {2018})}\BibitemShut
  {NoStop}%
\bibitem [{\citenamefont {Zhang}\ \emph {et~al.}(2008)\citenamefont {Zhang},
  \citenamefont {Genov}, \citenamefont {Wang}, \citenamefont {Liu},\ and\
  \citenamefont {Zhang}}]{zhang2008plasmon}%
  \BibitemOpen
  \bibfield  {author} {\bibinfo {author} {\bibfnamefont {S.}~\bibnamefont
  {Zhang}}, \bibinfo {author} {\bibfnamefont {D.~A.}\ \bibnamefont {Genov}},
  \bibinfo {author} {\bibfnamefont {Y.}~\bibnamefont {Wang}}, \bibinfo {author}
  {\bibfnamefont {M.}~\bibnamefont {Liu}}, \ and\ \bibinfo {author}
  {\bibfnamefont {X.}~\bibnamefont {Zhang}},\ }\href@noop {} {\bibfield
  {journal} {\bibinfo  {journal} {Phys. Rev. Lett.}\ }\textbf {\bibinfo
  {volume} {101}},\ \bibinfo {pages} {047401} (\bibinfo {year}
  {2008})}\BibitemShut {NoStop}%
\bibitem [{\citenamefont {Liu}\ \emph {et~al.}(2009)\citenamefont {Liu},
  \citenamefont {Langguth}, \citenamefont {Weiss}, \citenamefont {K{\"a}stel},
  \citenamefont {Fleischhauer}, \citenamefont {Pfau},\ and\ \citenamefont
  {Giessen}}]{liu2009plasmonic}%
  \BibitemOpen
  \bibfield  {author} {\bibinfo {author} {\bibfnamefont {N.}~\bibnamefont
  {Liu}}, \bibinfo {author} {\bibfnamefont {L.}~\bibnamefont {Langguth}},
  \bibinfo {author} {\bibfnamefont {T.}~\bibnamefont {Weiss}}, \bibinfo
  {author} {\bibfnamefont {J.}~\bibnamefont {K{\"a}stel}}, \bibinfo {author}
  {\bibfnamefont {M.}~\bibnamefont {Fleischhauer}}, \bibinfo {author}
  {\bibfnamefont {T.}~\bibnamefont {Pfau}}, \ and\ \bibinfo {author}
  {\bibfnamefont {H.}~\bibnamefont {Giessen}},\ }\href@noop {} {\bibfield
  {journal} {\bibinfo  {journal} {Nat. Mater.}\ }\textbf {\bibinfo {volume}
  {8}},\ \bibinfo {pages} {758} (\bibinfo {year} {2009})}\BibitemShut {NoStop}%
\bibitem [{\citenamefont {Singh}\ \emph {et~al.}(2011)\citenamefont {Singh},
  \citenamefont {Al-Naib}, \citenamefont {Koch},\ and\ \citenamefont
  {Zhang}}]{singh2011sharp}%
  \BibitemOpen
  \bibfield  {author} {\bibinfo {author} {\bibfnamefont {R.}~\bibnamefont
  {Singh}}, \bibinfo {author} {\bibfnamefont {I.~A.}\ \bibnamefont {Al-Naib}},
  \bibinfo {author} {\bibfnamefont {M.}~\bibnamefont {Koch}}, \ and\ \bibinfo
  {author} {\bibfnamefont {W.}~\bibnamefont {Zhang}},\ }\href@noop {}
  {\bibfield  {journal} {\bibinfo  {journal} {Opt. Express}\ }\textbf {\bibinfo
  {volume} {19}},\ \bibinfo {pages} {6312} (\bibinfo {year}
  {2011})}\BibitemShut {NoStop}%
\bibitem [{\citenamefont {Singh}\ \emph {et~al.}(2009)\citenamefont {Singh},
  \citenamefont {Rockstuhl}, \citenamefont {Lederer},\ and\ \citenamefont
  {Zhang}}]{singh2009coupling}%
  \BibitemOpen
  \bibfield  {author} {\bibinfo {author} {\bibfnamefont {R.}~\bibnamefont
  {Singh}}, \bibinfo {author} {\bibfnamefont {C.}~\bibnamefont {Rockstuhl}},
  \bibinfo {author} {\bibfnamefont {F.}~\bibnamefont {Lederer}}, \ and\
  \bibinfo {author} {\bibfnamefont {W.}~\bibnamefont {Zhang}},\ }\href@noop {}
  {\bibfield  {journal} {\bibinfo  {journal} {Phys. Rev. B}\ }\textbf {\bibinfo
  {volume} {79}},\ \bibinfo {pages} {085111} (\bibinfo {year}
  {2009})}\BibitemShut {NoStop}%
\bibitem [{\citenamefont {Liu}\ \emph {et~al.}(2012)\citenamefont {Liu},
  \citenamefont {Gu}, \citenamefont {Singh}, \citenamefont {Ma}, \citenamefont
  {Zhu}, \citenamefont {Tian}, \citenamefont {He}, \citenamefont {Han},\ and\
  \citenamefont {Zhang}}]{liu2012electromagnetically}%
  \BibitemOpen
  \bibfield  {author} {\bibinfo {author} {\bibfnamefont {X.}~\bibnamefont
  {Liu}}, \bibinfo {author} {\bibfnamefont {J.}~\bibnamefont {Gu}}, \bibinfo
  {author} {\bibfnamefont {R.}~\bibnamefont {Singh}}, \bibinfo {author}
  {\bibfnamefont {Y.}~\bibnamefont {Ma}}, \bibinfo {author} {\bibfnamefont
  {J.}~\bibnamefont {Zhu}}, \bibinfo {author} {\bibfnamefont {Z.}~\bibnamefont
  {Tian}}, \bibinfo {author} {\bibfnamefont {M.}~\bibnamefont {He}}, \bibinfo
  {author} {\bibfnamefont {J.}~\bibnamefont {Han}}, \ and\ \bibinfo {author}
  {\bibfnamefont {W.}~\bibnamefont {Zhang}},\ }\href@noop {} {\bibfield
  {journal} {\bibinfo  {journal} {Appl. Phys. Lett.}\ }\textbf {\bibinfo
  {volume} {100}},\ \bibinfo {pages} {131101} (\bibinfo {year}
  {2012})}\BibitemShut {NoStop}%
\bibitem [{\citenamefont {Zhu}\ \emph {et~al.}(2013)\citenamefont {Zhu},
  \citenamefont {Yang}, \citenamefont {Gu}, \citenamefont {Jiang},
  \citenamefont {Yue}, \citenamefont {Tian}, \citenamefont {Tonouchi},
  \citenamefont {Han},\ and\ \citenamefont {Zhang}}]{zhu2013broadband}%
  \BibitemOpen
  \bibfield  {author} {\bibinfo {author} {\bibfnamefont {Z.}~\bibnamefont
  {Zhu}}, \bibinfo {author} {\bibfnamefont {X.}~\bibnamefont {Yang}}, \bibinfo
  {author} {\bibfnamefont {J.}~\bibnamefont {Gu}}, \bibinfo {author}
  {\bibfnamefont {J.}~\bibnamefont {Jiang}}, \bibinfo {author} {\bibfnamefont
  {W.}~\bibnamefont {Yue}}, \bibinfo {author} {\bibfnamefont {Z.}~\bibnamefont
  {Tian}}, \bibinfo {author} {\bibfnamefont {M.}~\bibnamefont {Tonouchi}},
  \bibinfo {author} {\bibfnamefont {J.}~\bibnamefont {Han}}, \ and\ \bibinfo
  {author} {\bibfnamefont {W.}~\bibnamefont {Zhang}},\ }\href@noop {}
  {\bibfield  {journal} {\bibinfo  {journal} {Nanotechnology}\ }\textbf
  {\bibinfo {volume} {24}},\ \bibinfo {pages} {214003} (\bibinfo {year}
  {2013})}\BibitemShut {NoStop}%
\bibitem [{\citenamefont {Han}\ \emph {et~al.}(2014)\citenamefont {Han},
  \citenamefont {Singh}, \citenamefont {Cong},\ and\ \citenamefont
  {Yang}}]{han2014engineering}%
  \BibitemOpen
  \bibfield  {author} {\bibinfo {author} {\bibfnamefont {S.}~\bibnamefont
  {Han}}, \bibinfo {author} {\bibfnamefont {R.}~\bibnamefont {Singh}}, \bibinfo
  {author} {\bibfnamefont {L.}~\bibnamefont {Cong}}, \ and\ \bibinfo {author}
  {\bibfnamefont {H.}~\bibnamefont {Yang}},\ }\href@noop {} {\bibfield
  {journal} {\bibinfo  {journal} {J. Phys. D: Appl. Phys.}\ }\textbf {\bibinfo
  {volume} {48}},\ \bibinfo {pages} {035104} (\bibinfo {year}
  {2014})}\BibitemShut {NoStop}%
\bibitem [{\citenamefont {Gu}\ \emph {et~al.}(2012)\citenamefont {Gu},
  \citenamefont {Singh}, \citenamefont {Liu}, \citenamefont {Zhang},
  \citenamefont {Ma}, \citenamefont {Zhang}, \citenamefont {Maier},
  \citenamefont {Tian}, \citenamefont {Azad}, \citenamefont {Chen} \emph
  {et~al.}}]{gu2012active}%
  \BibitemOpen
  \bibfield  {author} {\bibinfo {author} {\bibfnamefont {J.}~\bibnamefont
  {Gu}}, \bibinfo {author} {\bibfnamefont {R.}~\bibnamefont {Singh}}, \bibinfo
  {author} {\bibfnamefont {X.}~\bibnamefont {Liu}}, \bibinfo {author}
  {\bibfnamefont {X.}~\bibnamefont {Zhang}}, \bibinfo {author} {\bibfnamefont
  {Y.}~\bibnamefont {Ma}}, \bibinfo {author} {\bibfnamefont {S.}~\bibnamefont
  {Zhang}}, \bibinfo {author} {\bibfnamefont {S.~A.}\ \bibnamefont {Maier}},
  \bibinfo {author} {\bibfnamefont {Z.}~\bibnamefont {Tian}}, \bibinfo {author}
  {\bibfnamefont {A.~K.}\ \bibnamefont {Azad}}, \bibinfo {author}
  {\bibfnamefont {H.-T.}\ \bibnamefont {Chen}},  \emph {et~al.},\ }\href@noop
  {} {\bibfield  {journal} {\bibinfo  {journal} {Nat. Commun.}\ }\textbf
  {\bibinfo {volume} {3}},\ \bibinfo {pages} {1151} (\bibinfo {year}
  {2012})}\BibitemShut {NoStop}%
\bibitem [{\citenamefont {Chatzakis}\ \emph {et~al.}(2012)\citenamefont
  {Chatzakis}, \citenamefont {Luo}, \citenamefont {Wang}, \citenamefont {Shen},
  \citenamefont {Koschny}, \citenamefont {Zhou},\ and\ \citenamefont
  {Soukoulis}}]{chatzakis2012reversible}%
  \BibitemOpen
  \bibfield  {author} {\bibinfo {author} {\bibfnamefont {I.}~\bibnamefont
  {Chatzakis}}, \bibinfo {author} {\bibfnamefont {L.}~\bibnamefont {Luo}},
  \bibinfo {author} {\bibfnamefont {J.}~\bibnamefont {Wang}}, \bibinfo {author}
  {\bibfnamefont {N.-H.}\ \bibnamefont {Shen}}, \bibinfo {author}
  {\bibfnamefont {T.}~\bibnamefont {Koschny}}, \bibinfo {author} {\bibfnamefont
  {J.}~\bibnamefont {Zhou}}, \ and\ \bibinfo {author} {\bibfnamefont
  {C.}~\bibnamefont {Soukoulis}},\ }\href@noop {} {\bibfield  {journal}
  {\bibinfo  {journal} {Phys. Rev. B}\ }\textbf {\bibinfo {volume} {86}},\
  \bibinfo {pages} {125110} (\bibinfo {year} {2012})}\BibitemShut {NoStop}%
\bibitem [{\citenamefont {Roy~Chowdhury}\ \emph {et~al.}(2013)\citenamefont
  {Roy~Chowdhury}, \citenamefont {Singh}, \citenamefont {Taylor}, \citenamefont
  {Chen},\ and\ \citenamefont {Azad}}]{roy2013ultrafast}%
  \BibitemOpen
  \bibfield  {author} {\bibinfo {author} {\bibfnamefont {D.}~\bibnamefont
  {Roy~Chowdhury}}, \bibinfo {author} {\bibfnamefont {R.}~\bibnamefont
  {Singh}}, \bibinfo {author} {\bibfnamefont {A.~J.}\ \bibnamefont {Taylor}},
  \bibinfo {author} {\bibfnamefont {H.-T.}\ \bibnamefont {Chen}}, \ and\
  \bibinfo {author} {\bibfnamefont {A.~K.}\ \bibnamefont {Azad}},\ }\href@noop
  {} {\bibfield  {journal} {\bibinfo  {journal} {Appl. Phys. Lett.}\ }\textbf
  {\bibinfo {volume} {102}},\ \bibinfo {pages} {011122} (\bibinfo {year}
  {2013})}\BibitemShut {NoStop}%
\bibitem [{\citenamefont {Cao}\ \emph {et~al.}(2013)\citenamefont {Cao},
  \citenamefont {Singh}, \citenamefont {Zhang}, \citenamefont {Han},
  \citenamefont {Tonouchi},\ and\ \citenamefont {Zhang}}]{cao2013plasmon}%
  \BibitemOpen
  \bibfield  {author} {\bibinfo {author} {\bibfnamefont {W.}~\bibnamefont
  {Cao}}, \bibinfo {author} {\bibfnamefont {R.}~\bibnamefont {Singh}}, \bibinfo
  {author} {\bibfnamefont {C.}~\bibnamefont {Zhang}}, \bibinfo {author}
  {\bibfnamefont {J.}~\bibnamefont {Han}}, \bibinfo {author} {\bibfnamefont
  {M.}~\bibnamefont {Tonouchi}}, \ and\ \bibinfo {author} {\bibfnamefont
  {W.}~\bibnamefont {Zhang}},\ }\href@noop {} {\bibfield  {journal} {\bibinfo
  {journal} {Appl. Phys. Lett.}\ }\textbf {\bibinfo {volume} {103}},\ \bibinfo
  {pages} {101106} (\bibinfo {year} {2013})}\BibitemShut {NoStop}%
\bibitem [{\citenamefont {Xu}\ \emph {et~al.}(2016)\citenamefont {Xu},
  \citenamefont {Su}, \citenamefont {Ouyang}, \citenamefont {Xu}, \citenamefont
  {Cao}, \citenamefont {Zhang}, \citenamefont {Li}, \citenamefont {Hu},
  \citenamefont {Gu}, \citenamefont {Tian} \emph {et~al.}}]{xu2016frequency}%
  \BibitemOpen
  \bibfield  {author} {\bibinfo {author} {\bibfnamefont {Q.}~\bibnamefont
  {Xu}}, \bibinfo {author} {\bibfnamefont {X.}~\bibnamefont {Su}}, \bibinfo
  {author} {\bibfnamefont {C.}~\bibnamefont {Ouyang}}, \bibinfo {author}
  {\bibfnamefont {N.}~\bibnamefont {Xu}}, \bibinfo {author} {\bibfnamefont
  {W.}~\bibnamefont {Cao}}, \bibinfo {author} {\bibfnamefont {Y.}~\bibnamefont
  {Zhang}}, \bibinfo {author} {\bibfnamefont {Q.}~\bibnamefont {Li}}, \bibinfo
  {author} {\bibfnamefont {C.}~\bibnamefont {Hu}}, \bibinfo {author}
  {\bibfnamefont {J.}~\bibnamefont {Gu}}, \bibinfo {author} {\bibfnamefont
  {Z.}~\bibnamefont {Tian}},  \emph {et~al.},\ }\href@noop {} {\bibfield
  {journal} {\bibinfo  {journal} {Opt. Lett.}\ }\textbf {\bibinfo {volume}
  {41}},\ \bibinfo {pages} {4562} (\bibinfo {year} {2016})}\BibitemShut
  {NoStop}%
\bibitem [{\citenamefont {Manjappa}\ \emph {et~al.}(2017)\citenamefont
  {Manjappa}, \citenamefont {Srivastava}, \citenamefont {Solanki},
  \citenamefont {Kumar}, \citenamefont {Sum},\ and\ \citenamefont
  {Singh}}]{manjappa2017hybrid}%
  \BibitemOpen
  \bibfield  {author} {\bibinfo {author} {\bibfnamefont {M.}~\bibnamefont
  {Manjappa}}, \bibinfo {author} {\bibfnamefont {Y.~K.}\ \bibnamefont
  {Srivastava}}, \bibinfo {author} {\bibfnamefont {A.}~\bibnamefont {Solanki}},
  \bibinfo {author} {\bibfnamefont {A.}~\bibnamefont {Kumar}}, \bibinfo
  {author} {\bibfnamefont {T.~C.}\ \bibnamefont {Sum}}, \ and\ \bibinfo
  {author} {\bibfnamefont {R.}~\bibnamefont {Singh}},\ }\href@noop {}
  {\bibfield  {journal} {\bibinfo  {journal} {Adv. Mater.}\ }\textbf {\bibinfo
  {volume} {29}} (\bibinfo {year} {2017})}\BibitemShut {NoStop}%
\bibitem [{\citenamefont {Zhao}\ \emph
  {et~al.}(2016{\natexlab{a}})\citenamefont {Zhao}, \citenamefont {Yuan},
  \citenamefont {Zhu},\ and\ \citenamefont {Yao}}]{zhao2016graphene}%
  \BibitemOpen
  \bibfield  {author} {\bibinfo {author} {\bibfnamefont {X.}~\bibnamefont
  {Zhao}}, \bibinfo {author} {\bibfnamefont {C.}~\bibnamefont {Yuan}}, \bibinfo
  {author} {\bibfnamefont {L.}~\bibnamefont {Zhu}}, \ and\ \bibinfo {author}
  {\bibfnamefont {J.}~\bibnamefont {Yao}},\ }\href@noop {} {\bibfield
  {journal} {\bibinfo  {journal} {Nanoscale}\ }\textbf {\bibinfo {volume}
  {8}},\ \bibinfo {pages} {15273} (\bibinfo {year}
  {2016}{\natexlab{a}})}\BibitemShut {NoStop}%
\bibitem [{\citenamefont {He}\ \emph {et~al.}(2016)\citenamefont {He},
  \citenamefont {Lin}, \citenamefont {Liu},\ and\ \citenamefont
  {Shi}}]{he2016terahertz}%
  \BibitemOpen
  \bibfield  {author} {\bibinfo {author} {\bibfnamefont {X.}~\bibnamefont
  {He}}, \bibinfo {author} {\bibfnamefont {F.}~\bibnamefont {Lin}}, \bibinfo
  {author} {\bibfnamefont {F.}~\bibnamefont {Liu}}, \ and\ \bibinfo {author}
  {\bibfnamefont {W.}~\bibnamefont {Shi}},\ }\href@noop {} {\bibfield
  {journal} {\bibinfo  {journal} {Nanotechnology}\ }\textbf {\bibinfo {volume}
  {27}},\ \bibinfo {pages} {485202} (\bibinfo {year} {2016})}\BibitemShut
  {NoStop}%
\bibitem [{\citenamefont {Luo}\ \emph {et~al.}(2016)\citenamefont {Luo},
  \citenamefont {Cai}, \citenamefont {Xiang}, \citenamefont {Wang},
  \citenamefont {Ren}, \citenamefont {Zhang},\ and\ \citenamefont
  {Xu}}]{luo2016flexible}%
  \BibitemOpen
  \bibfield  {author} {\bibinfo {author} {\bibfnamefont {W.}~\bibnamefont
  {Luo}}, \bibinfo {author} {\bibfnamefont {W.}~\bibnamefont {Cai}}, \bibinfo
  {author} {\bibfnamefont {Y.}~\bibnamefont {Xiang}}, \bibinfo {author}
  {\bibfnamefont {L.}~\bibnamefont {Wang}}, \bibinfo {author} {\bibfnamefont
  {M.}~\bibnamefont {Ren}}, \bibinfo {author} {\bibfnamefont {X.}~\bibnamefont
  {Zhang}}, \ and\ \bibinfo {author} {\bibfnamefont {J.}~\bibnamefont {Xu}},\
  }\href@noop {} {\bibfield  {journal} {\bibinfo  {journal} {Opt. Express}\
  }\textbf {\bibinfo {volume} {24}},\ \bibinfo {pages} {5784} (\bibinfo {year}
  {2016})}\BibitemShut {NoStop}%
\bibitem [{\citenamefont {Xia}\ \emph {et~al.}(2016)\citenamefont {Xia},
  \citenamefont {Zhai}, \citenamefont {Wang}, \citenamefont {Sun},
  \citenamefont {Liu},\ and\ \citenamefont {Wen}}]{xia2016dynamically}%
  \BibitemOpen
  \bibfield  {author} {\bibinfo {author} {\bibfnamefont {S.-X.}\ \bibnamefont
  {Xia}}, \bibinfo {author} {\bibfnamefont {X.}~\bibnamefont {Zhai}}, \bibinfo
  {author} {\bibfnamefont {L.-L.}\ \bibnamefont {Wang}}, \bibinfo {author}
  {\bibfnamefont {B.}~\bibnamefont {Sun}}, \bibinfo {author} {\bibfnamefont
  {J.-Q.}\ \bibnamefont {Liu}}, \ and\ \bibinfo {author} {\bibfnamefont
  {S.-C.}\ \bibnamefont {Wen}},\ }\href@noop {} {\bibfield  {journal} {\bibinfo
   {journal} {Opt. Express}\ }\textbf {\bibinfo {volume} {24}},\ \bibinfo
  {pages} {17886} (\bibinfo {year} {2016})}\BibitemShut {NoStop}%
\bibitem [{\citenamefont {Fu}\ \emph {et~al.}(2016)\citenamefont {Fu},
  \citenamefont {Zhai}, \citenamefont {Li}, \citenamefont {Xia},\ and\
  \citenamefont {Wang}}]{fu2016dynamically}%
  \BibitemOpen
  \bibfield  {author} {\bibinfo {author} {\bibfnamefont {G.-L.}\ \bibnamefont
  {Fu}}, \bibinfo {author} {\bibfnamefont {X.}~\bibnamefont {Zhai}}, \bibinfo
  {author} {\bibfnamefont {H.-J.}\ \bibnamefont {Li}}, \bibinfo {author}
  {\bibfnamefont {S.-X.}\ \bibnamefont {Xia}}, \ and\ \bibinfo {author}
  {\bibfnamefont {L.-L.}\ \bibnamefont {Wang}},\ }\href@noop {} {\bibfield
  {journal} {\bibinfo  {journal} {J. Opt.}\ }\textbf {\bibinfo {volume} {19}},\
  \bibinfo {pages} {015001} (\bibinfo {year} {2016})}\BibitemShut {NoStop}%
\bibitem [{\citenamefont {He}\ \emph {et~al.}(2018)\citenamefont {He},
  \citenamefont {Liu}, \citenamefont {Lin},\ and\ \citenamefont
  {Shi}}]{he2018graphene}%
  \BibitemOpen
  \bibfield  {author} {\bibinfo {author} {\bibfnamefont {X.}~\bibnamefont
  {He}}, \bibinfo {author} {\bibfnamefont {F.}~\bibnamefont {Liu}}, \bibinfo
  {author} {\bibfnamefont {F.}~\bibnamefont {Lin}}, \ and\ \bibinfo {author}
  {\bibfnamefont {W.}~\bibnamefont {Shi}},\ }\href@noop {} {\bibfield
  {journal} {\bibinfo  {journal} {Opt. Express}\ }\textbf {\bibinfo {volume}
  {26}},\ \bibinfo {pages} {9931} (\bibinfo {year} {2018})}\BibitemShut
  {NoStop}%
\bibitem [{\citenamefont {Li}\ \emph {et~al.}(2016)\citenamefont {Li},
  \citenamefont {Cong}, \citenamefont {Singh}, \citenamefont {Xu},
  \citenamefont {Cao}, \citenamefont {Zhang}, \citenamefont {Tian},
  \citenamefont {Du}, \citenamefont {Han},\ and\ \citenamefont
  {Zhang}}]{li2016monolayer}%
  \BibitemOpen
  \bibfield  {author} {\bibinfo {author} {\bibfnamefont {Q.}~\bibnamefont
  {Li}}, \bibinfo {author} {\bibfnamefont {L.}~\bibnamefont {Cong}}, \bibinfo
  {author} {\bibfnamefont {R.}~\bibnamefont {Singh}}, \bibinfo {author}
  {\bibfnamefont {N.}~\bibnamefont {Xu}}, \bibinfo {author} {\bibfnamefont
  {W.}~\bibnamefont {Cao}}, \bibinfo {author} {\bibfnamefont {X.}~\bibnamefont
  {Zhang}}, \bibinfo {author} {\bibfnamefont {Z.}~\bibnamefont {Tian}},
  \bibinfo {author} {\bibfnamefont {L.}~\bibnamefont {Du}}, \bibinfo {author}
  {\bibfnamefont {J.}~\bibnamefont {Han}}, \ and\ \bibinfo {author}
  {\bibfnamefont {W.}~\bibnamefont {Zhang}},\ }\href@noop {} {\bibfield
  {journal} {\bibinfo  {journal} {Nanoscale}\ }\textbf {\bibinfo {volume}
  {8}},\ \bibinfo {pages} {17278} (\bibinfo {year} {2016})}\BibitemShut
  {NoStop}%
\bibitem [{\citenamefont {Xiao}\ \emph {et~al.}(2017)\citenamefont {Xiao},
  \citenamefont {Wang}, \citenamefont {Jiang}, \citenamefont {Yan},
  \citenamefont {Cheng}, \citenamefont {Wang},\ and\ \citenamefont
  {Xu}}]{xiao2017strong}%
  \BibitemOpen
  \bibfield  {author} {\bibinfo {author} {\bibfnamefont {S.}~\bibnamefont
  {Xiao}}, \bibinfo {author} {\bibfnamefont {T.}~\bibnamefont {Wang}}, \bibinfo
  {author} {\bibfnamefont {X.}~\bibnamefont {Jiang}}, \bibinfo {author}
  {\bibfnamefont {X.}~\bibnamefont {Yan}}, \bibinfo {author} {\bibfnamefont
  {L.}~\bibnamefont {Cheng}}, \bibinfo {author} {\bibfnamefont
  {B.}~\bibnamefont {Wang}}, \ and\ \bibinfo {author} {\bibfnamefont
  {C.}~\bibnamefont {Xu}},\ }\href@noop {} {\bibfield  {journal} {\bibinfo
  {journal} {J. Phys. D: Appl. Phys.}\ }\textbf {\bibinfo {volume} {50}},\
  \bibinfo {pages} {195101} (\bibinfo {year} {2017})}\BibitemShut {NoStop}%
\bibitem [{\citenamefont {Chen}\ and\ \citenamefont
  {Fan}(2017)}]{chen2017study}%
  \BibitemOpen
  \bibfield  {author} {\bibinfo {author} {\bibfnamefont {X.}~\bibnamefont
  {Chen}}\ and\ \bibinfo {author} {\bibfnamefont {W.}~\bibnamefont {Fan}},\
  }\href@noop {} {\bibfield  {journal} {\bibinfo  {journal} {Opt. Lett.}\
  }\textbf {\bibinfo {volume} {42}},\ \bibinfo {pages} {2034} (\bibinfo {year}
  {2017})}\BibitemShut {NoStop}%
\bibitem [{\citenamefont {Ordal}\ \emph {et~al.}(1985)\citenamefont {Ordal},
  \citenamefont {Bell}, \citenamefont {Alexander}, \citenamefont {Long},\ and\
  \citenamefont {Querry}}]{ordal1985optical}%
  \BibitemOpen
  \bibfield  {author} {\bibinfo {author} {\bibfnamefont {M.~A.}\ \bibnamefont
  {Ordal}}, \bibinfo {author} {\bibfnamefont {R.~J.}\ \bibnamefont {Bell}},
  \bibinfo {author} {\bibfnamefont {R.~W.}\ \bibnamefont {Alexander}}, \bibinfo
  {author} {\bibfnamefont {L.~L.}\ \bibnamefont {Long}}, \ and\ \bibinfo
  {author} {\bibfnamefont {M.~R.}\ \bibnamefont {Querry}},\ }\href@noop {}
  {\bibfield  {journal} {\bibinfo  {journal} {Appl. Opt.}\ }\textbf {\bibinfo
  {volume} {24}},\ \bibinfo {pages} {4493} (\bibinfo {year}
  {1985})}\BibitemShut {NoStop}%
\bibitem [{\citenamefont {Zhang}\ \emph {et~al.}(2015)\citenamefont {Zhang},
  \citenamefont {Zhu}, \citenamefont {Liu}, \citenamefont {Yuan},\ and\
  \citenamefont {Qin}}]{zhang2015towards}%
  \BibitemOpen
  \bibfield  {author} {\bibinfo {author} {\bibfnamefont {J.}~\bibnamefont
  {Zhang}}, \bibinfo {author} {\bibfnamefont {Z.}~\bibnamefont {Zhu}}, \bibinfo
  {author} {\bibfnamefont {W.}~\bibnamefont {Liu}}, \bibinfo {author}
  {\bibfnamefont {X.}~\bibnamefont {Yuan}}, \ and\ \bibinfo {author}
  {\bibfnamefont {S.}~\bibnamefont {Qin}},\ }\href@noop {} {\bibfield
  {journal} {\bibinfo  {journal} {Nanoscale}\ }\textbf {\bibinfo {volume}
  {7}},\ \bibinfo {pages} {13530} (\bibinfo {year} {2015})}\BibitemShut
  {NoStop}%
\bibitem [{\citenamefont {Zhao}\ \emph
  {et~al.}(2016{\natexlab{b}})\citenamefont {Zhao}, \citenamefont {Zhang},
  \citenamefont {Zhu}, \citenamefont {Yuan},\ and\ \citenamefont
  {Qin}}]{zhao2016tunable}%
  \BibitemOpen
  \bibfield  {author} {\bibinfo {author} {\bibfnamefont {J.}~\bibnamefont
  {Zhao}}, \bibinfo {author} {\bibfnamefont {J.}~\bibnamefont {Zhang}},
  \bibinfo {author} {\bibfnamefont {Z.}~\bibnamefont {Zhu}}, \bibinfo {author}
  {\bibfnamefont {X.}~\bibnamefont {Yuan}}, \ and\ \bibinfo {author}
  {\bibfnamefont {S.}~\bibnamefont {Qin}},\ }\href@noop {} {\bibfield
  {journal} {\bibinfo  {journal} {J. Opt.}\ }\textbf {\bibinfo {volume} {18}},\
  \bibinfo {pages} {095001} (\bibinfo {year} {2016}{\natexlab{b}})}\BibitemShut
  {NoStop}%
\bibitem [{\citenamefont {Jnawali}\ \emph {et~al.}(2013)\citenamefont
  {Jnawali}, \citenamefont {Rao}, \citenamefont {Yan},\ and\ \citenamefont
  {Heinz}}]{jnawali2013observation}%
  \BibitemOpen
  \bibfield  {author} {\bibinfo {author} {\bibfnamefont {G.}~\bibnamefont
  {Jnawali}}, \bibinfo {author} {\bibfnamefont {Y.}~\bibnamefont {Rao}},
  \bibinfo {author} {\bibfnamefont {H.}~\bibnamefont {Yan}}, \ and\ \bibinfo
  {author} {\bibfnamefont {T.~F.}\ \bibnamefont {Heinz}},\ }\href@noop {}
  {\bibfield  {journal} {\bibinfo  {journal} {Nano Lett.}\ }\textbf {\bibinfo
  {volume} {13}},\ \bibinfo {pages} {524} (\bibinfo {year} {2013})}\BibitemShut
  {NoStop}%
\bibitem [{\citenamefont {Xiao}\ \emph {et~al.}(2018)\citenamefont {Xiao},
  \citenamefont {Wang}, \citenamefont {Liu}, \citenamefont {Yan}, \citenamefont
  {Li},\ and\ \citenamefont {Xu}}]{xiao2018active}%
  \BibitemOpen
  \bibfield  {author} {\bibinfo {author} {\bibfnamefont {S.}~\bibnamefont
  {Xiao}}, \bibinfo {author} {\bibfnamefont {T.}~\bibnamefont {Wang}}, \bibinfo
  {author} {\bibfnamefont {T.}~\bibnamefont {Liu}}, \bibinfo {author}
  {\bibfnamefont {X.}~\bibnamefont {Yan}}, \bibinfo {author} {\bibfnamefont
  {Z.}~\bibnamefont {Li}}, \ and\ \bibinfo {author} {\bibfnamefont
  {C.}~\bibnamefont {Xu}},\ }\href@noop {} {\bibfield  {journal} {\bibinfo
  {journal} {Carbon}\ }\textbf {\bibinfo {volume} {126}},\ \bibinfo {pages}
  {271} (\bibinfo {year} {2018})}\BibitemShut {NoStop}%
\bibitem [{\citenamefont {Lu}, \citenamefont {Liu},\ and\ \citenamefont
  {Mao}(2012)}]{lu2012plasmonic}%
  \BibitemOpen
  \bibfield  {author} {\bibinfo {author} {\bibfnamefont {H.}~\bibnamefont
  {Lu}}, \bibinfo {author} {\bibfnamefont {X.}~\bibnamefont {Liu}}, \ and\
  \bibinfo {author} {\bibfnamefont {D.}~\bibnamefont {Mao}},\ }\href@noop {}
  {\bibfield  {journal} {\bibinfo  {journal} {Phys. Rev. A}\ }\textbf {\bibinfo
  {volume} {85}},\ \bibinfo {pages} {053803} (\bibinfo {year}
  {2012})}\BibitemShut {NoStop}%
\end{thebibliography}

%

\end{document}